# A Lattice Calculation of the Heavy Quark Universal Form Factor

Jeffrey E. Mandula[a] and Michael C. Ogilvie[b]

[a]U.S. Department of Energy, Division of High Energy Physics, Washington, DC 20585, USA

[b]Department of Physics, Washington University, St. Louis, MO 63130, USA

A computation of the Isgur-Wise Universal Function using a lattice implementation of the heavy quark effective theory is described.

In this talk we review the lattice implementation of the heavy quark effective theory, describe its application to the computation of the universal heavy quark form factor, and show the first simulation results.

Two other calculations of this decay form factor treating the heavy quarks as ordinary quarks but with a small hopping constant, instead of using the lattice heavy quark theory are also being presented at this conference. [1]

The Isgur-Wise limit is the situation in which the mass of heavy quark is taken to be much larger than $\Lambda_{QCD}$, which is the characteristic energy of the strong interactions. [2] One separates the quark four-momentum into a fixed part proportional to the heavy mass and a residual dynamical part

$$P = Mv + p \qquad v^2 = 1 \qquad (1)$$

The unit four-vector $v$ is referred to as the classical velocity. In the infinite mass limit the quark propagator greatly simplifies:

$$\frac{1}{\gamma \cdot P - M} \to \frac{1 + \gamma \cdot v}{2 v \cdot p} \qquad (2)$$

The mass of the heavy quark drops out, and the Dirac structure of the propagator becomes simply a projector on positive energy states. This is, of course, the origin of the famous heavy quark spin-flavor symmetry. The heavy quark limit can be formulated as a field theory, which facilitates both non-perturbative analysis, the focus of this talk, and a systematic treatment of corrections in the inverse heavy quark mass. One constructs a field for a heavy quark of velocity $v$ by factoring the rapid phase proportional to the large mass from the canonical field,

$$h^{(v)}(x) = e^{-iMv \cdot x} \frac{1 + \gamma \cdot v}{2} \psi(x) \qquad (3)$$

There is an independent field for each classical velocity. [3] The Lagrangian for $h^{(v)}$,

$$\mathcal{L}^{(v)} = \bar{h}^{(v)}(x) \, iv \cdot D \, h^{(v)}(x) \qquad (4)$$

called the heavy quark effective theory, is the starting point of our analysis.

To incorporate the heavy quark effective theory into a lattice calculation, we Wick rotate it into Euclidean space. [4] The classical velocity, which is only a parameter, is complex when expressed in terms of Euclidean four-momenta:

$$v = (iv_0, \vec{v})$$
$$v_0 = \sqrt{1 + \vec{v}^2} \qquad (5)$$

The reduced propagator for the heavy quark satisfies the Euclidean space equation

$$-iv \cdot \partial \, \tilde{S}^{(v)}(x) = \delta(x) \qquad (6)$$

with the boundary condition

$$\tilde{S}^{(v)}(x) = 0 \qquad x_4 < 0 \qquad (7)$$

In the presence of gauge interactions, the derivative is replaced by a gauge covariant derivative. In order to most simply incorporate the boundary condition on the lattice, we represent the time derivative in the reduced Dirac equation by means of a forward difference, but use a symmetrical centered first difference for the space derivative. The lattice equation for the reduced propagator is:

$$v_0 [ U(x,x+\hat{t}) \tilde{S}^{(v)}(x+\hat{t},y) - \tilde{S}^{(v)}(x,y) ]$$
$$+ \sum_{\mu=1}^{3} \frac{-iv_\mu}{2} [U(x,x+\hat{\mu}) \tilde{S}^{(v)}(x+\hat{\mu},y) \quad (8)$$
$$- U(x,x-\hat{\mu}) \tilde{S}^{(v)}(x-\hat{\mu},y) ] = \delta(x,y)$$

This is solved by simple forward recursion; no iterative techniques are needed.

Even for non-zero $\vec{v}$, only trivial computation is needed to construct the heavy quark propagator. This makes it entirely feasible to compute three-point functions, and one may envision optimizing physical applications by a careful choice of heavy-light particle wave functions or replacing the simple first difference by an improved one which eliminates the leading lattice spacing corrections.

The lattice heavy quark theory has some unfamiliar features which are worth noting. The symmetrical first difference gives rise to doublers. However, unlike the usual fermion situation, their contributions automatically vanish in the zero spacing limit, without the necessity of adding a Wilson-like term. The reason is that they always occur in association with gluon modes whose momenta are at the edge of the Brillouin zone, and those modes have energies on the order of the inverse lattice spacing. This is easily seen in perturbation theory. [5] To leading order in the inverse heavy quark mass, there are no closed heavy quark loops.

Another feature of the lattice heavy quark theory is that for residual momenta oriented oppositely to the classical velocity, the propagator grows with Euclidean time. The reason is that the falloff is governed by the difference between the actual energy of a mode and that of the zero-residual-momentum mode, which is $\vec{v} \cdot \vec{p}/v_0$ in the heavy quark limit. When this negative, the falloff becomes growth. The gluon modes always compensate for this growth in simulations, but in perturbation theory, consistency requires an appropriate choice of contour in the Euclidean energy plane. [5,6]

With the lattice heavy quark theory described above, we are in a position to address the paradigmatic heavy quark process, $B \rightarrow D^* l \nu$, the weak decay of one meson containing a heavy quark into another.

$$B \rightarrow D l \nu \quad , \quad D^* l \nu$$
$$\langle D v' | \bar{c} \gamma_\mu (1 - i\gamma_5) b | B v \rangle \sim \xi(v \cdot v') \quad (9)$$

All the dynamical information is contained in the function $\xi$, known as the Isgur-Wise function, or the heavy quark universal form factor. It is a function only of the dot product of the classical momenta, and is normalized to unity for forward decay.

$$\xi(1) = 1 \quad (10)$$

The amplitude is contained in the reduced three-point function

$$\tilde{G}^{(v_B,v_D)}(x_0,z_0,y_0) \quad (11)$$
$$\sim \left\langle Tr \, s(y,x) \Gamma_D \tilde{S}^{(v_D)}(x,z) \Gamma_J \tilde{S}^{(v_B)}(z,y) \Gamma_B \right\rangle$$

The Isgur-Wise function may be extracted from the three-point correlation by dividing out the normalized single meson propagators. Better yet, we may take advantage of the normalization in the forward direction and use the three-point function at $v_D = v_B$ as normalization. The expression from which we extract the universal form factor is then

$$|\xi(v \cdot v')|^2 =$$
$$\lim_{x_0 \gg z_0 \gg y_0} \frac{\tilde{G}^{(v,v')}(x_0,y_0,z_0) \, \tilde{G}^{(v',v)}(x_0,y_0,z_0)}{\tilde{G}^{(v,v)}(x_0,y_0,z_0) \, \tilde{G}^{(v',v')}(x_0,y_0,z_0)} \quad (12)$$

This form has the advantage that the overall normalization of the three-point function drops out completely. Only those renormalizations that are classical velocity-dependent need be included.

We evaluated the three-point functions using the lattices made publicly available by Bernard and Soni. There were an ensemble of 16 lattices of size $16^3 \times 24$ with lattice coupling $\beta = 5.7$. The light quarks were represented by Wilson quark propagators with hopping constant $\kappa = .164$. The heavy quark propagators were evaluated for four values of each component of the initial and final quark classical velocity: 0, .25, .50, and .75. That is, there were $4^6 = 4096$ combinations of $v_B$ and $v_D$. Of course, the many that were lattice rotations of each other were averaged to improve statistics, but even so there were more than a hundred different values of $v_B \cdot v_D$, lying in the interval [1,1.822].

To improve the overlap of the meson wave function with the ground state, we applied multiple single-link smearing steps. Five steps gave statistically cleaner signals than none, but ten steps was not an improvement over five.

In order to project out the mesonic ground states from the three-point function, one either needs excellent wave functions or large Euclidean time separations between the slices on which the heavy meson is created, emits a weak current, and is detected. Unfortunately, the statistical noise in the simulation overwhelms the signal when those separations are as large as 4 units of Euclidean time. The largest separations that provide a signal are

$$x_4 - y_4 = 3 \qquad (13)$$
$$y_4 - z_4 = 3$$

At 2 units of Euclidean time separation, which corresponds to a greater contamination with higher mass states than at 3 units, the computed form factor is flatter. This presumably reflects Bjorken's sum rule, which states that the sum of the squares of the Isgur-Wise functions to all final states is 1. [7]

The results of the simulation together with a comparison to the data on $B \to D^* l \nu$ from both ARGUS and CLEO are shown in Fig. 1. [8] The agreement with the experimental data is rather spectacular. A simple quadratic fit to the simulation results gives for the slope at the origin

$$\xi'(1) = -.95 \qquad (14)$$

The strengths of the lattice heavy quark theory we have applied here are the same as those of the Isgur-Wise limit itself. By eliminating the heavy quark masses before undertaking a lattice analysis, only the residual four-momentum of the heavy quark is constrained by the inverse lattice spacing, and so there is in principle no restriction on $v_B \cdot v_D$.

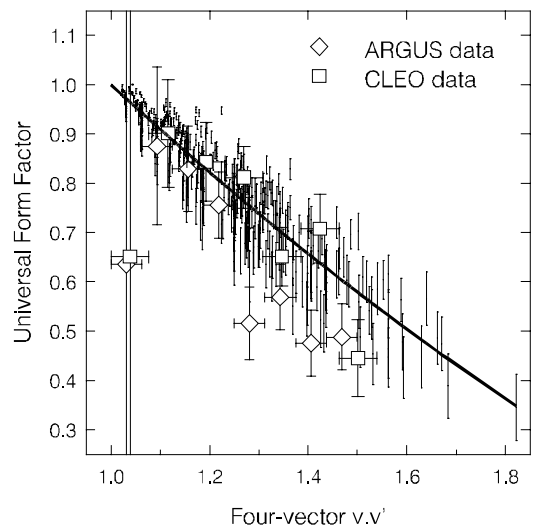

Figure 1. Lattice calculation of the Isgur-Wise form factor. Verticle bars are statistical errors for each simulation point. The solid curve is a quadratic fit to the simulation results.

The calculation described here can be improved by a more optimal choice of wave function, by including the velocity-dependent continuum [9] and lattice [6] renormalizations, and by the use of improved lattices and propagators from which the leading finite lattice spacing errors have been removed. We hope to have results of such a calculation by the next lattice meeting.